\documentclass[aps,prb,groupedaddress,showpacs]{revtex4}
\usepackage{graphicx}
\usepackage{helvet}
\usepackage{amsmath,amssymb}
\usepackage{bm}

\makeatletter
\newcommand{\Mtmdt}{[M(tmdt)$_{2}$]}
\newcommand{\mtmdt}{M(tmdt)$_{2}$}
\newcommand{\Nitmdt}{[Ni(tmdt)$_{2}$]}
\newcommand{\Pttmdt}{[Pt(tmdt)$_{2}$]}
\newcommand{\Autmdt}{[Au(tmdt)$_{2}$]}
\newcommand{\Cutmdt}{[Cu(tmdt)$_{2}$]}
\newcommand{\proton}{$^{1}$H}
\newcommand{\BC}{$^{13}$C}
\newcommand{\muB}{${\mu}_{\rm B}$}
\newcommand{\Tone}{$T_{1}^{-1}$}
\newcommand{\pTone}{$^{1}T_{1}^{-1}$}
\newcommand{\cTone}{$^{13}T_{1}^{-1}$}
\newcommand{\cToneT}{$(^{13}T_{1}T)^{-1}$}
\newcommand{\iTone}{$^{i}T_{1}^{-1}$}
\makeatother
\begin{document}

\title{Electronic States and Molecular Dynamics of Single-Component Molecular Conductor [M(tmdt)$_{2}$] (M=Ni, Pt) Studied by $^{13}$C and $^{1}$H NMR}
\author{Rina\ Takagi,\ Kazuya\ Miyagawa,\ Masahide\ Yoshimura,\ Hiro\ Gangi,\ and\ Kazushi\ Kanoda}
\affiliation{Department\ of\ Applied Physics,\ University\ of\ Tokyo,\ Bunkyo\ City,\ Tokyo,\ 113-8656,\ Japan}
\author{Biao\ Zhou,\ Yuki\ Idobata,\ and\ Akiko\ Kobayashi}
\affiliation{Department\ of\ Chemistry,\ College\ of\ Humanities\ and\ Sciences,\ Nihon\ University,\ Setagaya\ City,\ Tokyo,\ 156-8550,\ Japan} 
 

\begin{abstract}
 The molecular conductors  {\Mtmdt} (M=Ni, Pt) consisting of single molecular species are investigated with $^{13}$C NMR and $^{1}$H NMR. The temperature dependences of $^{13}$C NMR shift and relaxation rate provide microscopic evidences for the metallic nature with appreciable electron correlations. Both compounds exhibit an anomalous frequency-dependent enhancement in $^{1}$H nuclear spin-lattice relaxation rate in a wide temperature range. These observations signify the presence of extraordinary molecular motions with low energy excitations.

\end{abstract}

\maketitle

\section{Introduction}
 Orbital degrees of freedom are among ingredients that make the properties of materials fertile, in particular in strongly correlated electron systems. A family of single-component molecular ${\pi}-d$ systems, {\Mtmdt}, are multi-orbital correlated electron systems, where M and tmdt stand for metallic ion and organic ligand, trimethylenetetrathiafulvalenedithiolate, respectively. \cite{1,2,3} First-principles band-structure calculations point to the quasi-degenerate feature of molecular orbitals, which is a key to the emergence of metallic states from a single molecular species unlike the charge-transfer type of conductors. The molecular orbitals located near the Fermi level consist of $dp{\sigma}$-orbital, centered on M and extended to the neighboring sulfur atoms, and $p{\pi}$ orbitals extended over tmdt. \cite{4,5} The energy-level difference between the $dp{\sigma}$ and $p{\pi}$ orbitals depends on M and its variation gives different ground states even among isostructural compounds; a one-dimensional antiferromagnetic Mott insulator for M = Cu, an antiferromagnetic metal for M = Au and paramagnetic metals for M = Ni and Pt. \cite{6,7,8,9} 

 M=Au and Cu compounds, where the $dp{\sigma}$ and $p{\pi}$ orbitals are energetically closer to each other than in M=Ni and Pt salts, exhibit antiferromagnetic transitions at 110 K and 13 K, respectively \cite{10,11}. {\Cutmdt} shows the temperature profile of NMR relaxation rate characteristic of one-dimensional antiferromagnetic Heisenberg chains in the paramagnetic state and a contracted moment of $0.2-0.5$ {\muB} in the ordered state. \cite{11} These results support that the spins are on the $dp{\sigma}$-orbitals, which form quasi-one-dimensional chains according to the molecular orbital calculation \cite{5,9}. As for {\Autmdt}, the analysis of the {\proton} NMR spectra found a sizable moment of $0.7-1.2$ {\muB}/tmdt \cite{10}, which is considerably larger than expected in a spin-density-wave (SDW) state suggested by the ab-initio band calculations.\cite{12} The antiferromagnetic transition in {\Autmdt} has not yet been fully understood; the multi-orbital character might be pertinent to the magnetism.

 In {\Nitmdt} and {\Pttmdt}, on the other hand, $dp{\sigma}$ and $p{\pi}$ orbitals are well separated in energy and the Fermi level is located in conduction bands constructed solely from $p{\pi}$ orbitals. \cite{4,8,13} The magnetic susceptibility shows the Pauli-paramagnetic temperature dependence in both compounds. \cite{6,8} As for {\Nitmdt}, semi-metallic Fermi surfaces have been experimentally indicated, consistent with first-principles calculations.\cite{7} Although these experimental and theoretical studies show that {\Nitmdt} is (semi-)metallic, the further in-depth characterization of the electronic and molecular states has remained to be seen. In the present work, we carried out {\BC} NMR and {\proton} NMR measurements on the two compounds at ambient pressure. Both compounds show typical metallic behavior in {\BC} NMR relaxation rate, that is, Korringa relation with an appreciable correlation-induced enhancement factor, approximately above 50 K, below which nuclear relaxations due to impurity spins become apparent. {\proton} NMR relaxation rate shows unexpected temperature and frequency dependences, which are discussed in terms of peculiar molecular motions with low-lying excitations in {\Mtmdt}.

\section{EXPERIMENTAL}
 The {\BC} and {\proton} NMR experiments were performed for powdered crystals of {\Mtmdt} (M=Ni, Pt) with a standard NMR spectrometer at temperatures between 1.7 K and 300 K. The NMR spectra were obtained by the fast Fourier transformation of echo signals. The nuclear spin-lattice relaxation curves were obtained from the recovery of the echo intensity following saturation comb pulses.

 The {\BC} NMR measurements were carried out for the {\BC} isotope-enriched samples. (The enriched carbon sites are indicated in the inset of Fig. 9.) We used the spin echo pulse sequences of $({\pi}/2)_{x}-({\pi})_{x}$ to obtain NMR spectra and relaxation curves. As the origin of the NMR shift, we referred to the {\BC} NMR line of TMS (tetramethylsilane). To obtain the {\proton}-NMR signals, we employed the so-called solid-echo pulse sequence of $({\pi}/2)_{x}-({\pi}/2)_{y}$, where x and y stand for the axes in the rotational frame. \cite{14} In order to examine the field/frequency dependence of the NMR spectra and relaxation rate, we performed NMR measurements at several applied fields up to 11.5 Tesla.

\section{RESULTS AND DISCUSSIONS}
\subsection{{\BC} NMR}
 Figure 1 shows {\BC} NMR spectra for {\Nitmdt} and {\Pttmdt} at 252 K and 253 K, respectively. The lineshape has asymmetry typical of powder patterns of the Knight shift with uniaxial symmetry. This feature is consistent with the theoretical consequence that the orbital responsible for the conduction band is $p{\pi}$ orbitals, the spins on which generate anisotropic hyperfine fields at {\BC} sites mainly through the on-site $2p_{\rm z}$ dipolar coupling. As temperature is decreased, however, the line is gradually broadened with the asymmetry less appreciable, very probably due to progressive contribution of inhomogeneity to the lineshape. The isotropic component of the Knight shift, ${\delta}_{\rm iso}$, is determined by subtracting the isotropic component of chemical shift, ${\delta}^{\rm chem}_{\rm iso}$, from the first moment (the center of gravity) of the measured NMR spectra. As the value of ${\delta}^{\rm chem}_{\rm iso}$, we employed the separately measured chemical shift of the non-magnetic [Zn(tmdt)$_{2}$] with monovalent tmdt as in M=Ni and Pt; that is ${\delta}^{\rm chem}_{\rm iso} = 126$ ppm. The temperature dependence of ${\delta}_{\rm iso}$ and the line width, $W$, defined by the square root of the second moment of the spectra are shown in Figs. 2(a) and 2(b). The errors in the estimate of ${\delta}_{\rm iso}$ and $W$ are within ${\pm} 5$ ppm in the whole temperature range. The spin susceptibility, ${\chi}$, shown in Fig. 2(c) was obtained by subtracting the core diamagnetism and the Curie contribution from the static susceptibility measured with a SQUID magnetometer. The impurity-spin concentrations determined from the Curie contribution are 0.6 ${\%}$ for M=Pt and 0.8 ${\%}$ for M=Ni although NMR data described later suggest that the former contains more impurity-spins than the latter. A small hump apparent at low temperatures is likely an artifact of the Curie-term subtraction.

 The ${\chi}$ exhibits monotonous and weak temperature dependence in both compounds. The magnitude of ${\chi}$ is approximately 1.8 times larger in M = Ni than in M = Pt. The first-principles calculations, however, show almost the same densities of states near the Fermi level for the two compounds, which correspond to $1.4{\times}10^{-4}$ emu/mol in ${\chi}$ close to the experimental value for Pt compound. We have no clear explanation on this discrepancy but it might be due to electron correlation, as discussed later. The ${\delta}_{\rm iso}$ monotonically decreases with temperature as ${\chi}$ does and they roughly scale to each other for both systems although the temperature variation of ${\delta}_{\rm iso}$ is more remarkable than that of ${\chi}$. As mentioned in Introduction, the conduction bands of M = Ni and Pt salts consist of the $p{\pi}$ orbitals solely.\cite{4,8,13} The NMR shift is proportional to the product of the hyperfine coupling constant at {\BC} site and the local spin susceptibility. Comparing the data in Figs. 2 (a) and 2(c), the isotropic part of the hyperfine coupling constant, that is the ratio of ${\delta}_{\rm iso}$ to ${\chi}$, is $a_{\rm iso}$ = 3.9 and 5.1 kOe/({\muB} tmdt) for M=Ni and Pt when evaluated by averaging the values above 200 K, indicating that the Mulliken population of the $p_{\rm z}$ orbital responsible for the hyperfine coupling is somewhat different between M=Pt and Ni. The linewidth $W$ increases gradually as temperature is decreased in both compounds, suggesting the development of local-field inhomogeneity.
 
 Nuclear spin-lattice relaxation time, $T_{1}$, is usually defined as a characteristic time of the exponential decay of nuclear magnetization, $M({\infty})-M(t) {\propto} {\exp}[-(t/T_{1})]$, for nuclei with spin $I=1/2$ such as {\BC} and {\proton} nuclei. In case of anisotropic hyperfine coupling as in the present material, however, the recovery of {\BC} nuclear magnetization in powdered samples does not obey a single-exponential function of time; so, we defined $T_{1}$ by fitting the so-called stretched-exponential function, $M({\infty})-M(t){\propto} {\exp}[-(t/T_{1})^{\beta}]$, to the relaxation curves. The standard errors in the estimate of $T_{1}$ and ${\beta}$ are mostly less than 6 ${\%}$ for both M=Pt and Ni compounds. The fitting exponent, ${\beta}$, which characterizes the degree of distribution in $T_{1}$, is plotted in the inset of Fig. 3; ${\beta} = 0.6-0.7$ for M = Pt is smaller than ${\beta} = 0.7-0.8$ for M = Ni, suggesting an additional distribution in $T_{1}$ in {\Pttmdt}. Thus determined relaxation rate divided by temperature, {\cToneT}, is shown in Fig. 3. Overall, {\cToneT} is roughly constant in 100 $-$ 300 K for both compounds, which is conventional behavior of metallic systems. To be precise, {\cToneT} for {\Nitmdt} shows a positive temperature coefficient as ${\delta}_{\rm iso}$ does. The magnitude of {\cToneT} for {\Nitmdt} is larger than for {\Pttmdt} in accordance with their relative magnitude of ${\chi}$ and ${\delta}_{\rm iso}$. At lower temperatures below 100 K, {\cToneT} shows a gradual increase, which is more prominent in {\Pttmdt} than {\Nitmdt}. This can be caused by paramagnetic impurities like the linewidth behavior.
 
 The spin fluctuations as a manifestation of electron correlation in a paramagnetic metal is characterized by the so-called Korringa ratio, $K({\alpha})$, which is the relaxation enhancement factor in the Korringa relation and is given, in the simple case of isotropic hyperfine coupling, as
\begin{equation}\label{eq1}
\frac{1}{T_{1}T}= K({\alpha})\frac{4{\pi}k_{\rm B}}{{\hbar}}\left( \frac{{\gamma}_{\rm e}}{{\gamma}_{\rm n}}\right)^{2}{\delta}_{\rm iso}^{2}.
\end{equation}
In case of anisotropic hyperfine coupling as in the present compounds, however, Eq. (1) is modified to a form with field angle as a parameter. Further, when a sample is a powder, relaxation curves with angle-dependent {\Tone} are summed up into a non-single exponential function. As the volume-average of distributed {\Tone} is given by the relaxation rate of the initial slope, {\iTone}, Eq. (1) is extended to a form expressed by $^{i}T_{1}$. We re-analyzed the relaxation curves to determine $^{i}T_{1}$ and confirmed that the temperature dependence of $(^{i}T_{1}T)^{-1}$ (the inset of Fig. 4) is similar to that shown in Fig. 3. The chemical structure surrounding the {\BC} site in tmdt is nearly the same as that in ET, where the hyperfine field at the {\BC} site mainly consist of an isotropic core-polarization field and a uniaxially symmetric dipole field from the on-site $p_{\rm z}$ orbital\cite{15}. Thus, the Knight shifts of the present systems are characterized by an isotropic part, ${\delta}_{\rm iso}$, and an anisotropic part, ${\delta}_{\rm aniso}$. In this case, Eq. (1) is modified to\cite{15}
\begin{equation}
\frac{1}{^{i}T_{1}T}= K({\alpha})\frac{4{\pi}k_{B}}{{\hbar}}\left(\frac{{\gamma}_{\rm e}}{{\gamma}_{\rm n}}\right)^{2}({\delta}_{\rm iso}^{2} + 2{\delta}_{\rm aniso}^{2}).
\end{equation}

 To evaluate $K({\alpha})$, one has to know the values of ${\delta}_{\rm aniso}$, which can be extracted from the analysis of the spectra. Using the chemical shift tensor of non-magnetic and isovalent [Zn(tmdt)$_{2}$] \cite{16}, the total-shift tensor is given by
\begin{equation}
\begin{pmatrix}
\delta_{\rm xx} &0 &0\\
0 &{\delta}_{\rm yy} &0\\
0 &0 &{\delta}_{\rm zz}
\end{pmatrix}
= 126 + 
\begin{pmatrix}
47 &0 &0\\
0 &-4.4 &0\\
0 &0 &-42.8
\end{pmatrix}
+ {\delta}_{\rm iso} + {\delta}_{\rm aniso}
\begin{pmatrix}
-1 &0 &0\\
0 &-1 &0\\
0 &0 &2
\end{pmatrix}
,
\end{equation}
where x and y axes are in the molecular plane and z axis is perpendicular to them. The former two terms express the chemical shifts and the latter two terms do the Knight shifts. The total shift for a field directing to a spherical coordination, (${\theta}$, ${\phi}$), is given by ${\delta}({\theta}, {\phi}) = {\delta}_{\rm xx}{\sin}^{2}{\theta} {\cos}^{2}{\phi} + {\delta}_{\rm yy}{\sin}^{2}{\theta}{\sin}^{2}{\phi} + {\delta}_{\rm zz}{\cos}^{2}{\theta}$. Its powder-distribution, $f({\delta})$, has parameters, ${\delta}_{\rm iso}$ and ${\delta}_{\rm aniso}$, which are obtained by fitting $f({\delta})$ to the experimental spectra. In reality, however, inevitable inhomogeneity imposes additional broadening on the spectra. As seen in Fig. 2(b), the temperature dependence of $W$ is not scaled to that of ${\delta}_{\rm iso}$, indicating a sizable contribution of inhomogeneous broadening to the lineshape at low temperatures. Thus, in the analysis for extracting the values of ${\delta}_{\rm aniso}$, we incorporated the inhomogeneous broadening by convoluting $f({\delta})$ with a Lorentzian function in the following form,
\begin{eqnarray}
F(\delta)&=&\int f(\omega) \dfrac{{\Delta}}{({\omega}-{\delta})^{2}+{\Delta}^{2}} d{\omega}\nonumber \\
&=& \iint \dfrac{{\Delta}}{({\delta}_{\rm xx}{\sin}^{2}{\theta}{\cos}^{2}{\phi} + {\delta}_{\rm yy}{\sin}^{2}{\theta}{\sin}^{2}{\phi} + {\delta}_{\rm zz}{\cos}^{2}{\theta}-{\delta})^2 + {\Delta}^2} {\sin}{\theta} d{\theta}d{\phi},
\end{eqnarray}
where ${\Delta}$ characterizes the inhomogeneous width and is assumed to have a form of ${\Delta}^2 = {\Delta}_{0}^{2}+a{\omega}^{2}$ with the second term to express the width dependent on the shift. The form of Eq. (4) including four parameters (${\delta}_{\rm iso}$, ${\delta}_{\rm aniso}$, ${\Delta}_{0}$ and $a$) is fitted to the measured spectra. At high temperatures (above 100 K for M=Ni and above 200 K for M=Pt), the fitting was successful, as seen in Fig. 1, with reasonable sets of parameter values. At lower temperatures, however, the fitting parameter values are scattered and unsettled, as expected because the lineshape becomes more isotropic at low temperatures due to the dominant contribution of the inhomogeneous broadening and ${\delta}_{\rm aniso}$ is not resolved in the fitting. Nevertheless, considering that the ratio, ${\delta}_{\rm aniso}$/${\delta}_{\rm iso}$, should be temperature-independent because the form of the hyperfine tensor should not change under temperature variation, we use the reliable value of ${\delta}_{\rm aniso}$/${\delta}_{\rm iso}$ that is determined at high temperatures; ${\delta}_{\rm aniso}$/${\delta}_{\rm iso}$ = 0.78 for M=Ni and ${\delta}_{\rm aniso}$/${\delta}_{\rm iso}$ = 0.81 for M=Pt. (Thus, the hyperfine coupling constants are ($a_{\rm iso}$, $a_{\rm aniso}$) = (3.9, 3.0) and (5.1, 4.1) in unit of kOe/({\muB} tmdt) for M=Ni and Pt, respectively.) Using fitting the lines for ${\delta}_{\rm iso}$ shown in Fig. 2(a) and the ${\delta}_{\rm aniso}$ values determined from the above ratios along with the curves fitting the $(^{i}T_{1}T)^{-1}$ data shown in the inset of Fig. 4, we evaluate $K({\alpha})$ at temperatures for $T>100$ K, in which NMR parameters are much less affected by inhomogeneity. As shown in Fig. 4, $K({\alpha})$ is estimated at $3-5$ for M=Ni and $3.5-7$ for M=Pt, indicating appreciable antiferromagnetic fluctuations. For reference, a quarter-filled charge-transfer salt,${\theta}$-(ET)$_{2}$I$_{3}$, shows $K({\alpha})= 2-3$\cite{17} and a half-filled system situated on the border of Mott transition\cite{18}, ${\kappa}$-(ET)$_{2}$Cu[N(CN)$_{2}$]Br, shows $K({\alpha})$=8\cite{15,19,20}. Thus, the present materials are addressed as systems with non-negligible electron correlation.

 Figure 5 shows the temperature dependence of nuclear spin-spin relaxation rate, $T_{2}^{-1}$, of {\Pttmdt}, which is determined by fitting a Gaussian function to a spin-echo decay curve, as shown in the inset, and therefore is denoted by $T_{2{\rm G}}^{-1}$ hereafter. In general, $T_{2{\rm G}}^{-1}$ is governed by the nuclear dipolar interactions, which are temperature-independent. Actually, $T_{2{\rm G}}^{-1}$ is nearly constant in temperatures from 2 K to 300 K. Slow dynamics at the order of ${\sim}$kHz or lower, if any, would contribute to the $T_{2}^{-1}$ relaxation with a temperature dependence specific to the dynamics. Besides, enhanced antiferromagnetic fluctuations would also enhance $T_{2{\rm G}}^{-1}$ through the RKKY-like indirect interactions between nuclear spins. However, there are no such signatures in the present results. 

\subsection{$^{1}$H NMR}
 Unlike {\BC} sites, the hyperfine coupling of {\proton} nuclear spins with conduction electrons is so small that the Knight shift is below the level of detection and therefore the spectral position is not changed against temperature variation. {\proton} NMR spectral shape is nearly the same for the two compounds and temperature-insensitive; a spectrum of {\Nitmdt} at 1.8 K is shown in the inset of Fig. 6. Figure 6 shows the temperature dependence of {\proton} NMR line width evaluated by the square root of the second moment of {\proton} NMR spectra of {\Nitmdt}. The value of the line width, approximately 17 kHz at room temperature, is reasonably explained by the {\proton} nuclear-dipole interactions in the trimethylene group.
 
 The temperature dependence of {\proton}-NMR relaxation rate, {\pTone}, measured at several frequencies is shown in Fig. 7. $^{1}T_{1}$ is defined by fitting the stretched-exponential function $M({\infty})-M(t) {\propto} {\exp}[-(t/T_{1})^{\beta}]$, to the experimental relaxation curves, which were non-single-exponential functions in the whole temperature region. The fitting exponent ${\beta}$ falls in a range between 0.7 and 0.8 and gradually decreases with temperature. The possible origin of the distribution in {\pTone} is discussed later. The temperature dependence of {\pTone} is totally distinct from that of {\cTone}. Unlike {\cTone}, the so-called Korringa relation, {\Tone}${\propto} T$, does not hold at all in {\pTone} for both compounds. Instead, anomalous peak structures and pronounced frequency dependences are clearly seen. {\pTone} remarkably increases as the frequency is decreased, implying that the fluctuations would be more enhanced at further lower frequencies than the NMR Larmor frequencies of ${\sim}$MHz. Such slow and frequency-dependent fluctuations that persist in a wide temperature range is not attributable to the conventional spin dynamics in a paramagnetic metal. Such frequency-dependent peak structures in {\pTone} appear due to the molecular motions as observed in ET compounds such as ${\beta}$-(ET)$_{2}$I$_{3}$,\cite{22} ${\kappa}$-(ET)$_{2}$Cu(NCS)$_{2}$\cite{23} and ${\kappa}$-(ET)$_{2}$Cu[N(CN)$_{2}$]Cl,\cite{24} where the terminal ethylene in the ET molecule has bi-stable configurations, as shown in Fig. 8 (a). The conformational change is thermally activated to give a peak structure in {\pTone} against temperature variation when the characteristic frequency of the motions, which decreases with temperature, pass through the NMR probe frequency, ${\omega}$.

 The NMR relaxation due to such molecular dynamics is caused through the coupling between the molecular motion and the nuclear dipolar field, and described by the Bloembergen-Purcell-Pound (BPP) relation,\cite{25}
\begin{equation}
\frac{1}{T_{1}} = C \left( \frac{{\tau}_{\rm C}}{1 + {\omega}^{2}{\tau}_{\rm C}^{2}} + 4\frac{{\tau}_{\rm C}}{1 + 4{\omega}^{2}{\tau}_{\rm C}^{2}} \right),
\end{equation}
where ${\tau}_{\rm C}$ is the correlation time of the molecular motion that varies with temperature and $C$ is a coupling constant between the nuclear relaxation and the molecular dynamics. If the motion is of such a type that it is activated over an energy barrier, ${\tau}_{\rm C}$ is expressed in a form of
\begin{equation}
{\tau}_{\rm C} = {\tau}_{0} {\exp}(T_{\rm g}/T), 
\end{equation}
where $T_{\rm g}$ is the excitation energy for the conformation change and ${\tau}_{0}$ is an attempt frequency. The {\pTone} in the ET compounds mentioned above is well described by Eqs. (5) and (6); e.g., $T_{\rm g}$ and ${\tau}_{0}$ of ${\beta}$-(ET)$_{2}$I$_{3}$ were estimated at $T_{\rm g}$ ${\simeq} 2000$ K and ${\tau}_{0}$ ${\simeq} 1{\times}10^{-11}$ sec and similar values are reported for other ET systems as well\cite{22,23,24}. Turning to the present systems, however, the skeleton of the trimethylene is flat and fits well to the molecular plane of {\mtmdt} (see Fig. 8 (b)); so it is unlikely that the structural dynamics of the trimethylene group contributes to {\pTone} in {\Nitmdt} and {\Pttmdt} as in ET molecules.
 
 To advance this argument, we measured {\pTone} for a powdered sample of neutral HMTTF molecules, which contain trimethylenes (see the insets of Fig. 9). Since HMTTF is a band insulator, {\proton} nuclear spins should relax mostly by the fluctuation of the nuclear dipolar field due to the molecular motions, if any. Figure 9 shows {\pTone} of HMTTF together with that of {\Nitmdt} measured at 156 MHz. It is obvious that {\pTone} of HMTTF is negligibly small and differs in temperature dependence, compared with {\pTone} of {\Nitmdt}, indicating that the molecular motions in question are not in the trimethylene. It is also evident that the low-temperature anomaly is not due to the trimethylene’s motion. Yet, they do not contain molecules which have tunneling rotation with small excitation energies as in methyl groups in TMTSF compounds.\cite{23}

 We attempt to characterize the frequency-dependent peak formations in {\pTone} vs. temperature in {\Nitmdt} and {\Pttmdt} in terms of the BPP type formula, Eq. (5). Although {\pTone} ought to include the contribution of the conducting electrons in the form of {\pTone}${\propto} T$, this contribution is overwhelmed by the supposed motional contribution, as is obvious in Fig. 7. Equation (5) gives the peaks in {\pTone} at ${\omega}{\tau}_{\rm C} = 0.616$, which determines the value of ${\tau}_{\rm C}$ at the peak temperature $T_{\rm P}$ (denoted as ${\tau}_{\rm P}$ hereafter) for each NMR probe frequency, ${\omega}$. An Arrhenius plot of ${\tau}_{\rm P}$ vs. $T_{\rm P}^{-1}$ is shown in the inset of Fig. 7(a). The data points are not on a straight line, suggesting that the motions are not of an activation type as expressed by Eq. (6). Furthermore, the frequency dependence of {\pTone} is not in line with the BPP behavior in that {\pTone} is frequency-insensitive at high frequencies; {\pTone} at 156 MHz and 340 MHz for {\Nitmdt} makes little difference and therefore may be in the high frequency limit. The similar situation is seen at 156 MHz and 256 MHz for {\Pttmdt}. On the other hand, Eq. (5) is reduced, in the low temperature regime of ${\omega} {\gg} {\tau}_{\rm C}^{-1}$, to a form of {\Tone}$=2C/{\omega}^{2}{\tau}_{\rm C}$ with a strong ${\omega}$-dependence, which contradicts the observation shown in Fig. 7. A conceivable situation in {\Mtmdt} is that a frequency-independent contribution to {\pTone}, which comes from a separate dynamics with the correlation time much faster than the NMR probe frequencies, is superposed on the frequency-dependent contribution. On this assumption, we subtracted the frequency-independent part, namely {\pTone} at 156 MHz, from {\pTone} at lower frequencies to make the BPP analysis solely for the frequency-dependent part, which is shown in Fig. 10. The inset shows the Arrhenius plot of ${\tau}_{\rm P}$ vs. $T_{\rm P}^{-1}$, which yields nominal values of $T_{g} =640$ K and ${\tau}_{0} = 4.7{\times}10^{-11}$ sec.  The temperature dependences of {\pTone} calculated with the use of the $T_{\rm g}$ and ${\tau}_{0}$ values are shown by solid curves in Fig. 10, where the optimized (dimensionless) fitting parameter $C$ is 175. As seen in the figure, however, there are sizable discrepancies between the experimental data and the calculations, in particular, in the low-temperature side of $T_{\rm P}$. Equation (5) assumes that the fluctuations are constant in amplitude but is variable in correlation time ${\tau}_{\rm C}$, which is further assumed in Eq. (6) to follow an exponential temperature dependence characteristic of a thermal activation over an energy barrier. The discrepancies indicate the molecular motions in question have different features from the assumptions which Eqs. (5) and (6) are based on.

 We find a hint to this puzzle in the first-principles calculations of the molecular orbital energy with respect to the dihedral angles (${\theta}$) between two tmdt ligands across the metallic ion.\cite{4} For M = Ni and Pt, {\mtmdt} is most stable at ${\theta}=0^{\circ}$; namely, two tmdt’s in a molecule are co-planar. For M = Cu and Zn, however, two tmdt’s are twisted. This suggests that the twisting degrees of freedom inherent in {\mtmdt} may be easily activated. Actually, the ${\theta}$-dependence of the molecular orbital energy for M = Ni and Pt is so flat around ${\theta}=0^{\circ}$, as seen in Fig. 4 of Ref. [4] and in Ref. [25]. Assuming that the tmdt is a rigid planar body with a moment of inertia and approximating the orbital energy vs. ${\theta}$ curve in a quadratic form around ${\theta}=0^{\circ}$ \cite{4, 26}, the natural frequency of the twisting vibration is estimated at ca. $1{\times}10^{10}$ Hz for both salts. This value corresponding to $5{\times}10^{-5}$ eV ${\simeq} 0.6$ K means nearly continuous excitations in the temperature scale studied here; the twisting angles are estimated at ${\pm}5$ degree for {\Nitmdt} and ${\pm}3$ degree for {\Pttmdt} at 100 K by referring to the calculated curves.\cite{4,26} Because the twisting motions do not associate with activation over an energy barrier as in ET molecules, the characteristic time of the motions is unlikely to vary in an activation manner. In solids, inter-molecular interactions make the local motions collective. However, it is an open question how the observed frequency and temperature dependences of {\pTone} is explained by the tmdt twisting motions specific to this family of materials.

 It is noted that $T_{2{\rm G}}^{-1}$ in {\BC} NMR does not show enhancement indicative of motions at the order of ${\sim}$kHz or lower. Two origins are conceivable. One is that the molecular motion is not strong at the {\BC} sites on the twisting axis. The other is the small hyperfine coupling of {\BC} nuclear spins with the motions. Unlike {\proton}, the local field fluctuations at {\BC} sites by the nuclear motions are through the temporal modulation of anisotropic total shift. [There should be no motional relaxation in case of isotropic shift.] It is likely that the motional fluctuations of the anisotropic part in the total shift are not enough to give a sizable contribution to $T_{2{\rm G}}^{-1}$ compared with the temperature-independent nuclear-dipolar contribution.

 Finally, we comment on low-temperature data well below 50 K. The frequency dependence of {\pTone} is distinct from that at higher temperatures, as seen in Fig. 7. For low frequencies, {\pTone} tends to a constant value in the low-temperature limit. For increased frequency, however, {\pTone} becomes to show a drop-off. As mentioned in the analysis of the {\BC} NMR shift, our NMR data suggest the effect of impurity spins present in the samples. Based on the analyses in Appendix, we conclude that the relaxation arising from the impurity spins becomes apparent at low temperatures where the intrinsic nuclear relaxations due to the conducting electrons and molecular dynamics become inappreciable. At high temperatures, the high frequency limit of {\pTone} in {\proton} NMR likely trace the impurity contributions (see Appendix). As for {\BC} NMR, {\cTone} due to impurity spins is overwhelmed by the Korringa contribution at high temperatures above 100 K (see Appendix).

\section{CONCLUSION}
 We have investigated the single-component molecular conductors, {\Nitmdt} and {\Pttmdt} by {\BC} NMR and {\proton} NMR measurements, which probed the electronic states and molecular dynamics. The {\BC} Knight shift and nuclear spin-lattice relaxation rate have proved the paramagnetic metallic states for both compounds through the observation of the Korringa relation. Further analysis gives the Korringa ratio, $K({\alpha})$, which indicates appreciable antiferromagnetic fluctuations in both compounds; that is, electron correlation is not negligible. The temperature and frequency dependences of the {\proton} NMR relaxation rate reveal unusual molecular dynamics presumably coming from the tmdt’s twisting degrees of freedom or some novel type of molecular motions specific to this family of materials.

\section{ACKNOWLEDGEMENTS}
 The authors thank S. Ishibashi and T. Itou for valuable discussions.

 This work was supported by JSPS Grant-in-Aids for Scientific Research (S) (Grant No. 25220709), for Challenging Exploratory Research (Grant No. 24654101), and for JSPS Fellows (Grant No. 13J03087), and MEXT Grant-in-Aids for Scientific Research on Innovative Areas (Grant No. 20110003) and for Strategic Research Base Development Program for Private Universities (2009) (Grant No. S0901022).

\section*{Appendix}
 In Figs. 11 (a) and (b), {\pTone} is plotted as a function of $T/H$. The data fall into a universal curve for $T$ below 11 K and $H$ above 3.7 T for both compounds, far better than plotted as a function of $T$ shown in Figs. 7 (c) and (d). Figure 11(c) shows the ${\Delta}${\cTone}, which is the low-temperature deviation of {\cTone} from the extrapolation of the Korringa relation holding at higher temperatures (100 $-$ 280 K), {\cTone} $= 7.6{\times}10^{-3}$ T for M=Pt and {\cTone} $= 9.5{\times}10^{-3}$ T for M=Ni. The ${\Delta}${\cTone} also appears to have similar functional dependence for $T$ below 7 K and $H$ above 5.0 T for M=Pt and $T$ below 6 K for M=Ni. Such behavior indicates that nuclear spin-lattice relaxation at low temperatures is governed by the fluctuation of the local field coupled with applied magnetic field, not by molecular dynamics. This type of relaxation scaling has been observed in intermetallic compounds containing paramagnetic impurities \cite{27,28}. Direct dipolar interactions between impurity spins and nuclear spins can cause a nuclear spin-lattice relaxation through the longitudinal and transverse interactions. In addition, in case of metals, nuclear spins are indirectly coupled with impurity spins through conduction electrons; that is, the RKKY interaction, which causes nuclear relaxation via two channels known as the Benoit-de Gennes-Silhoutte (BGS) mechanism\cite{29} and the Giovannini-Heeger (GH) mechanism\cite{30}. We found the experimental universal curve of {\pTone} well described by a function of $dB_{\rm S}(x)/dx$, as shown in Figs. 11 (a) and (b), where $B_{\rm S}(x)$ represents the Brillouin function of spin $S$ and the parameter $x = gS{\mu}_{\rm B}H/k_{\rm B}T$, with $S=1/2$ and $g=2$. The relaxation rate arising from the longitudinal dipolar interaction follows a function of $dB_{\rm S}(x)/dx$, whereas the three other mechanisms have $B_{\rm S}(x)/x$ dependence.\cite{28} Thus, {\pTone} at low temperatures is likely caused by the longitudinal dipolar interactions with impurity spins. Small upward deviations of the data from the theoretical curve for large values of $T/H$ are attributable to the Korringa contribution. As for ${\Delta}${\cTone}, however, the data are best described by a comparably weighted sum of $dB_{\rm S}(x)/dx$ and $B_{\rm S}(x)/x$. This suggests that other interactions with impurity spins contribute to the {\BC} nuclear relaxation along with the longitudinal dipolar interaction. At higher temperatures, {\cTone} shows the Korringa relation. Because {\Tone} arising from the RKKY interaction is channeled through the Korringa mechanism, {\cTone} at low temperatures can have appreciable contributions from the BGS and GH relaxation, compared to {\pTone}.
 
 Figures 11(a) and (b) show that, at low magnetic fields below ca. 3 T, {\pTone} deviates from the universal $T/H$ scaling curve, which should be followed when the free impurity spins are in the Zeeman splitting states. This deviation suggests that impurity spins have interactions of the order of 3 T (equivalent to 0.3 meV) with certain degrees of freedom other than magnetic field, which might be $J$ coupling with conduction electrons. Alternatively, it may be the low-temperature remnant of the anomalous molecular dynamics.

\newpage
\begin{figure}
\begin{center}
\includegraphics[width=8cm,keepaspectratio]{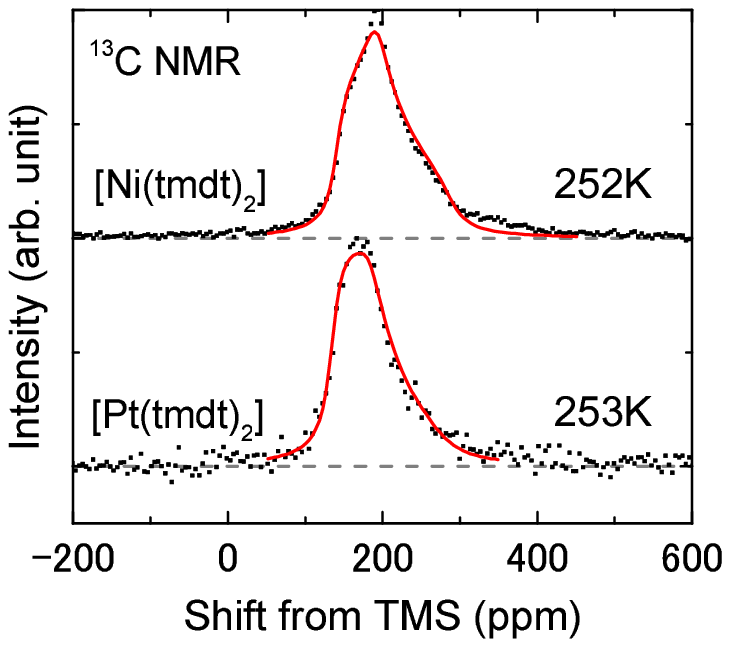}
\end{center}
\caption{(Color online) $^{13}$C NMR spectra of [Ni(tmdt)$_{2}$] and [Pt(tmdt)$_{2}$] at 252 K and 253 K, respectively, and fitting curves based on Eq. (4).}
\label{Fig1}
\end{figure}

\begin{figure}
\begin{center}
\includegraphics[width=7cm,keepaspectratio]{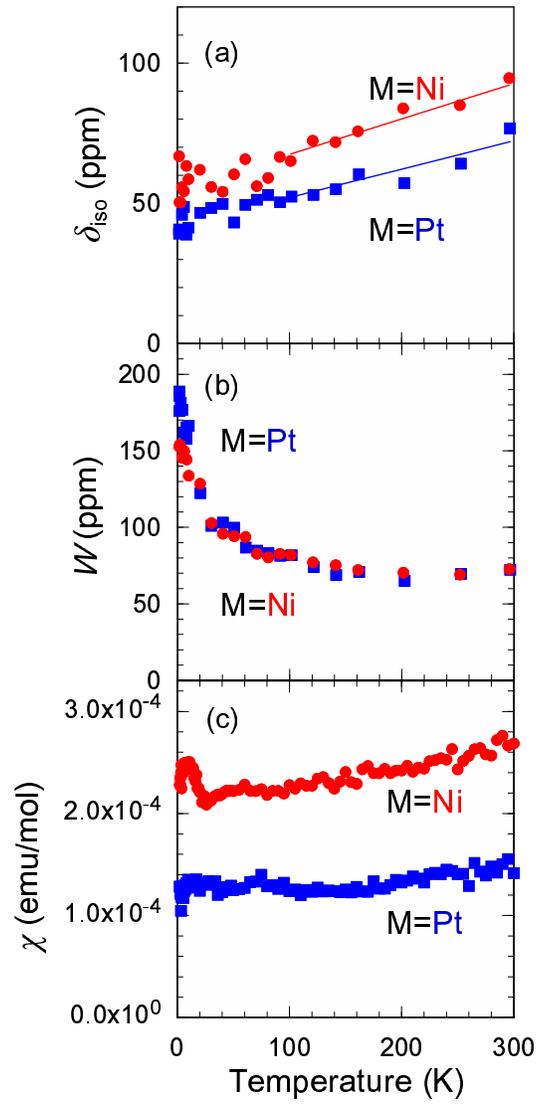}
\end{center}
\caption{(Color online) Temperature dependences of (a) isotropic components of $^{13}$C NMR Knight shift, (b) the line width of $^{13}$C NMR spectra, and (c) spin susceptibility for [Ni(tmdt)$_{2}$] and [Pt(tmdt)$_{2}$]. The shift origin of (a) is ${\delta}^{\rm chem}_{\rm iso}=126$ ppm. The lines in (a) are the linear-fitting to the raw data for $T > 100$ K, which are used in evaluating the Korringa ratio, $K({\alpha})$ (see text).}
\label{Fig2}
\end{figure}

\begin{figure}
\begin{center}
\includegraphics[width=8.6cm,keepaspectratio]{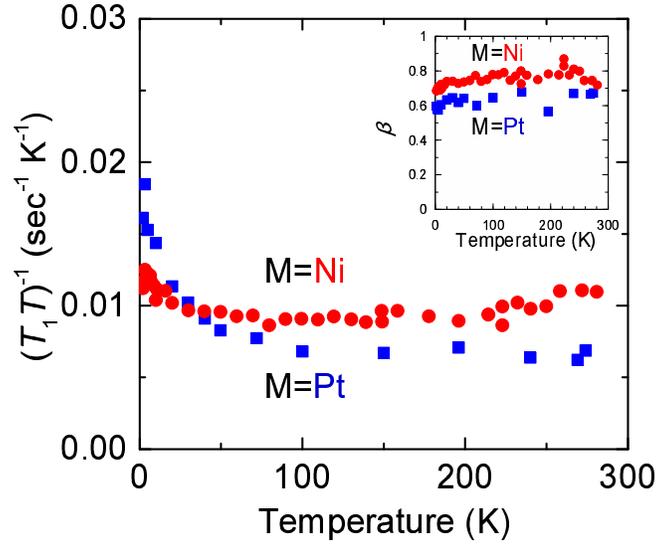}
\end{center}
\caption{(Color online) Temperature dependences of $^{13}$C NMR relaxation rate, defined by fitting the stretched-exponential function, divided by temperature. The inset shows the temperature dependence of the fitting exponent, ${\beta}$.}
\label{Fig3}
\end{figure}

\begin{figure}
\begin{center}
\includegraphics[width=8.6cm,keepaspectratio]{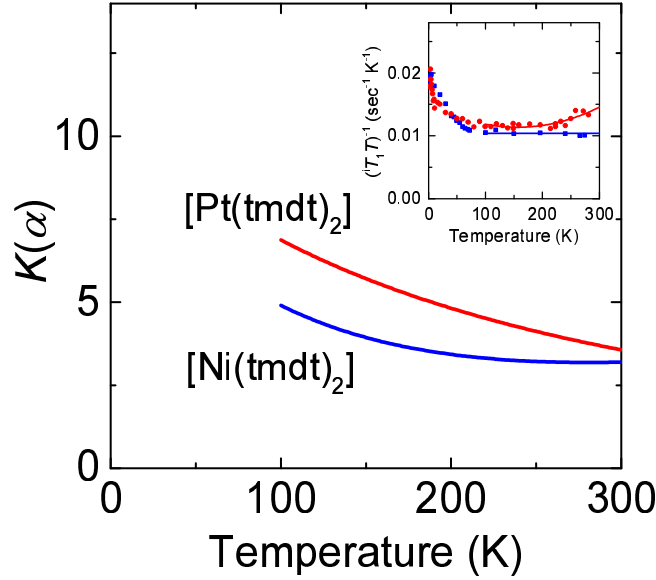}
\end{center}
\caption{(Color online) Temperature dependence of the NMR relaxation enhancement factor (so-called Korringa ratio, $K({\alpha})$) evaluated by Eq. (2). For the values of ${\delta}_{\rm iso}$, ${\delta}_{\rm aniso}$ and $(^{i}T_{1}T)^{-1}$ in the evaluation, the fitting curves to the ${\delta}_{\rm iso}$ data (Fig. 2(a)), the relations of ${\delta}_{\rm aniso}$ = 0.78${\delta}_{\rm iso}$ for M=Ni and ${\delta}_{\rm aniso}$ =0.81${\delta}_{\rm iso}$ for M=Pt (see text) and the fitting curves to the $(^{i}T_{1}T)^{-1}$ data (Inset) are used, respectively.}
\label{Fig4}
\end{figure}

\begin{figure}
\begin{center}
\includegraphics[width=8.6cm,keepaspectratio]{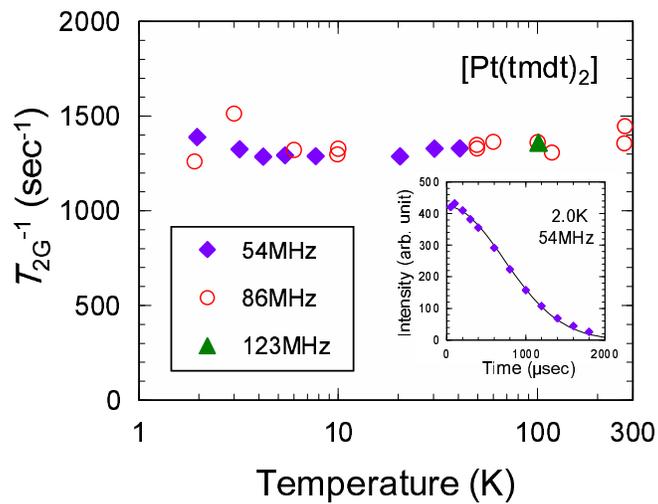}
\end{center}
\caption{ (Color online) Temperature dependence of $T_{2 \rm{G}}^{-1}$ of [Pt(tmdt)$_{2}$]. The inset shows a typical spin-echo decay with a fitting curve of a Gaussian function.}
\label{Fig5}
\end{figure}

\begin{figure}
\begin{center}
\includegraphics[width=8.6cm,keepaspectratio]{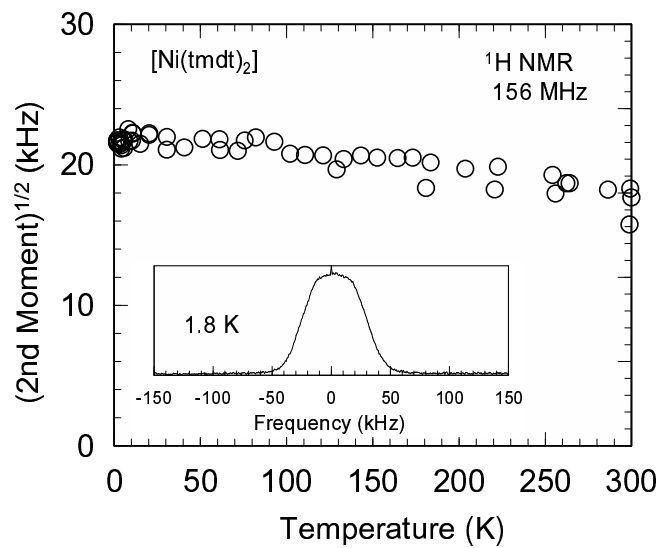}
\end{center}
\caption{Temperature dependence of the square root of the second moment of the $^{1}$H NMR spectra for [Ni(tmdt)$_{2}$]. The inset is the typical spectrum for [Ni(tmdt)$_{2}$] measured at 1.8 K.}
\label{Fig6}
\end{figure}

\begin{figure}
\begin{center}
\includegraphics[width=16cm,keepaspectratio]{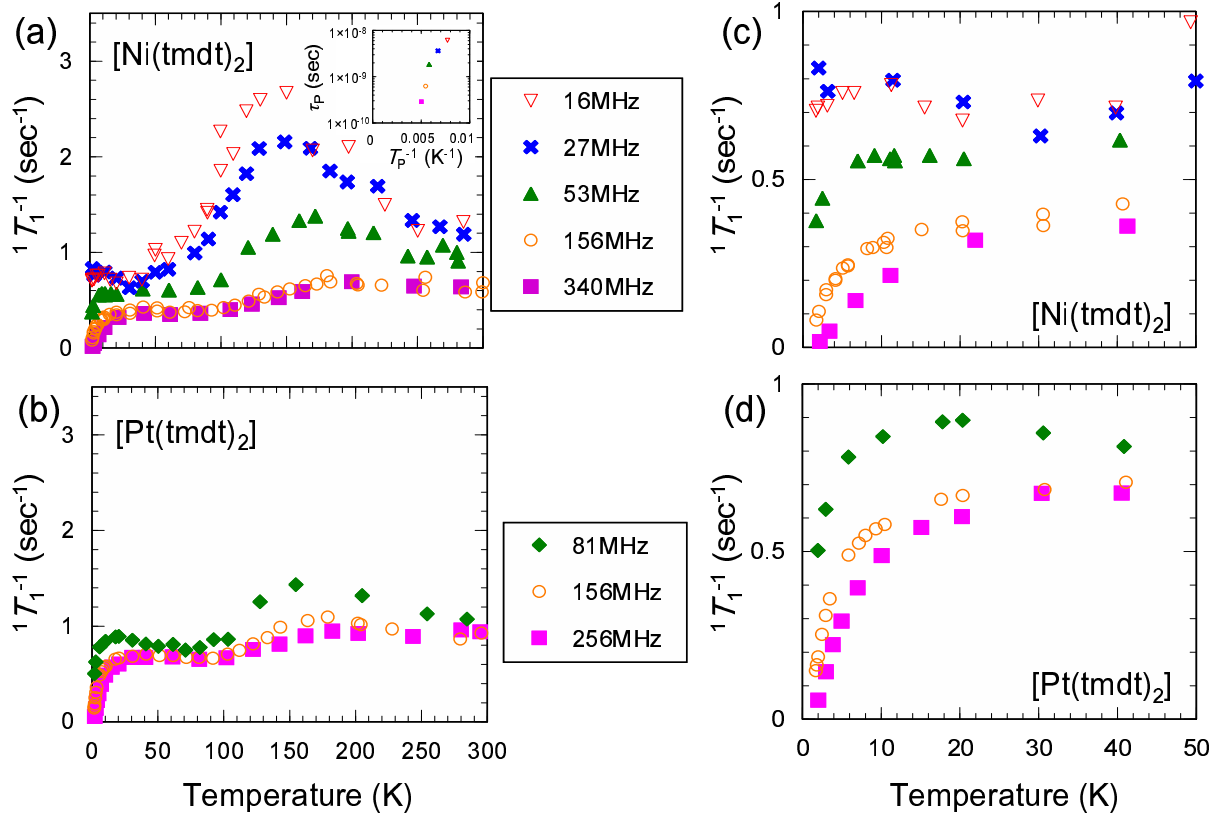}
\end{center}
\caption{(Color online) Temperature dependences of {\pTone} for {\Nitmdt} (a) and {\Pttmdt} (b). (c) and (d) show {\pTone} below 50 K, respectively. The inset in (a) shows ${\tau}_{\rm P}$ vs. $T_{\rm P}^{-1}$ of each NMR frequencies.}
\label{Fig7}
\end{figure}

\begin{figure}
\begin{center}
\includegraphics[width=14cm,keepaspectratio]{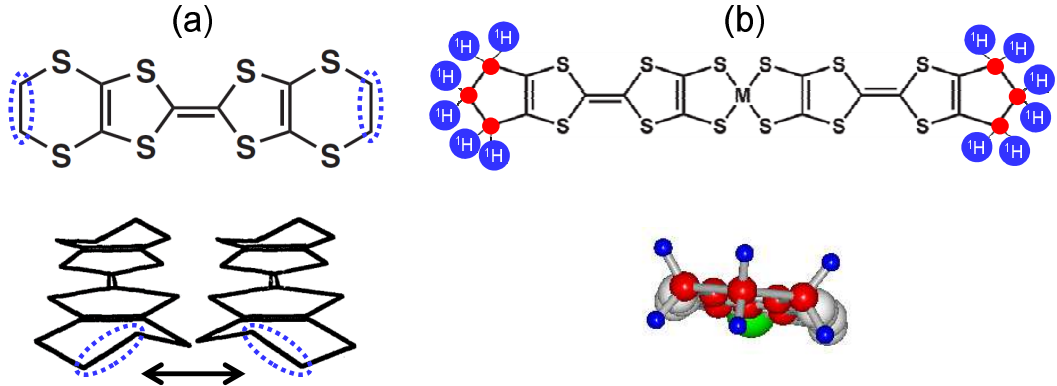}
\end{center}
\caption{(Color online) Molecular structure of (a) ET and (b) M(tmdt)$_{2}$. The bottoms are viewed from a direction along the long axis of the molecules. Ethylenes of the ET molecule are indicated with dotted ellipses in (a).}
\label{Fig8}
\end{figure}

\begin{figure}
\begin{center}
\includegraphics[width=8.6cm,keepaspectratio]{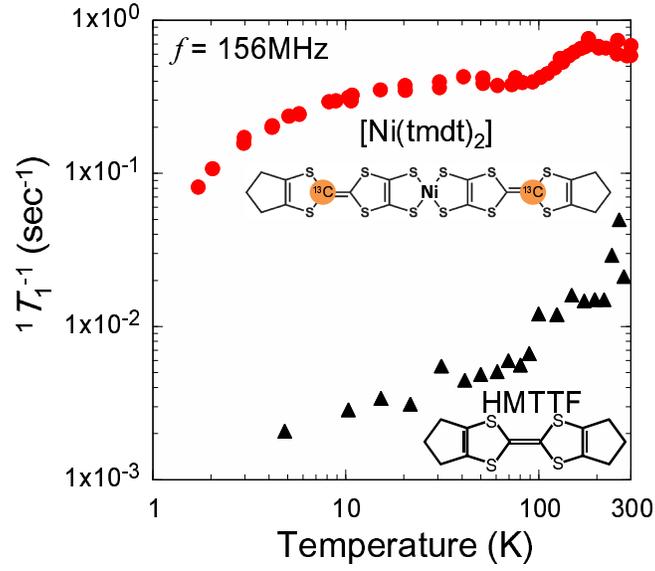}
\end{center}
\caption{(Color online) {\pTone} of {\Nitmdt} and HMTTF neutral solid. Both molecular structures show trimethylenes in the terminals.}
\label{Fig9}
\end{figure}

\begin{figure}
\begin{center}
\includegraphics[width=8.6cm,keepaspectratio]{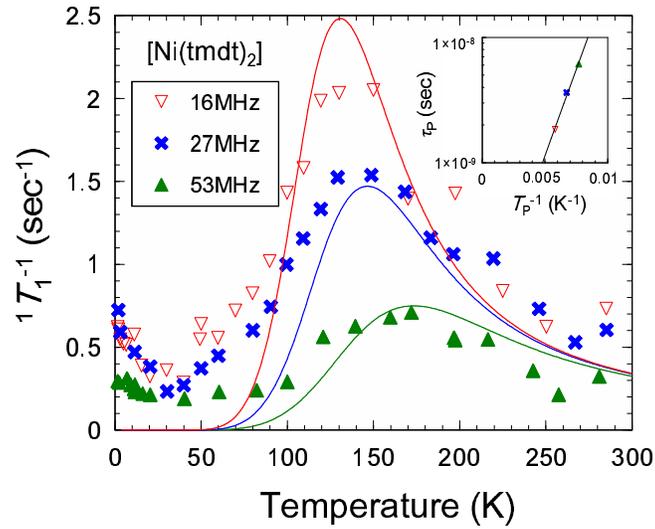}
\end{center}
\caption{(Color online) Temperature dependences of {\pTone} of 16 MHz, 27 MHz and 53 MHz subtracted that of 156 MHz for [Ni(tmdt)$_{2}$]. The solid curves are fitting ones at each frequency with BPP relation using Eq. (5) and a set of $T_{\rm g}=640$ K and ${\tau}_{0}=4.7{\times}10^{-11}$ sec evaluated the activation plot of $T_{\rm P}$ vs. ${\tau}_{\rm P}$ as shown in the inset.}
\label{Fig10}
\end{figure}

\begin{figure}
\begin{center}
\includegraphics[width=8.6cm,keepaspectratio]{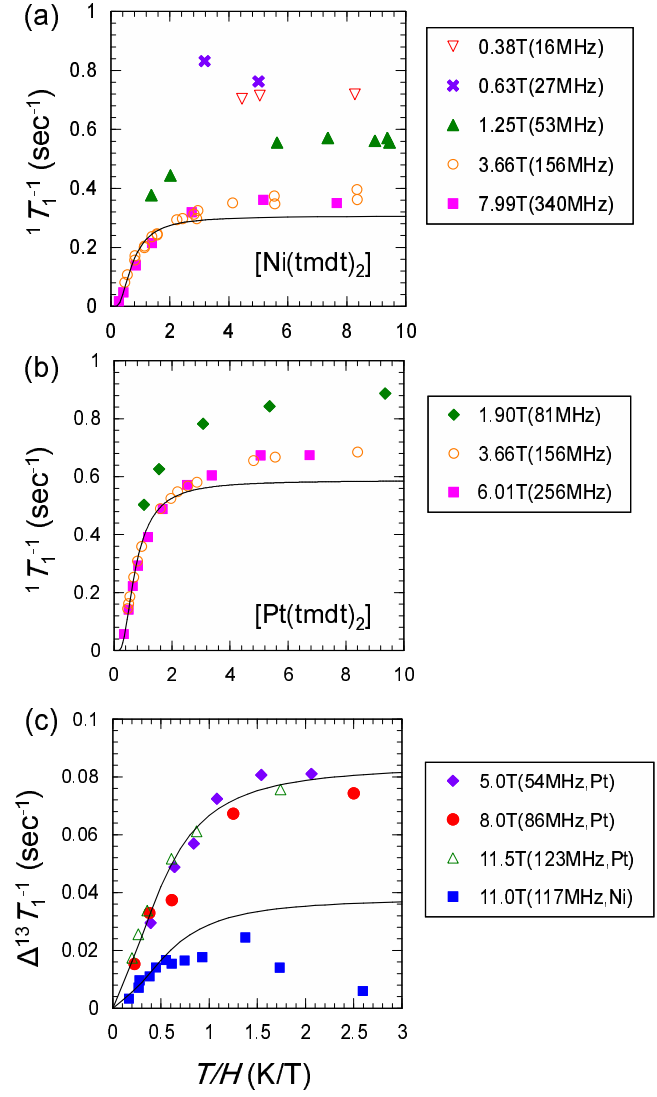}
\end{center}
\caption{(Color online) {\pTone} plotted against temperature divided by external field for (a) [Ni(tmdt)$_{2}$] and (b) [Pt(tmdt)$_{2}$]. (c) Deviation of {\cTone} from the Korringa relation holding at high temperatures plotted against temperature divided by external field for [Ni(tmdt)$_{2}$] and [Pt(tmdt)$_{2}$]. Solid curves are the fitting results with the function $dB_{\rm S}(x)/dx$ ($x=gS{\mu}_{\rm B}H/k_{\rm B}T$) for (a) and (b), and with the combination of the functions proportional to $dB_{\rm S}(x)/dx$ and $B_{\rm S}(x)/x$ for (c).}
\label{FigA1}
\end{figure}

\end{document}